\documentclass[aps,floats,prb,showpacs,twocolumn]{revtex4}
\usepackage{amsfonts,amsmath}
\usepackage{bm}
\usepackage{graphicx}
\usepackage{latexsym}
\usepackage{multirow}

\begin{document}

\title{Unconventional quantum criticality in the kicked rotor}

\author{Jiao Wang$^1$, Chushun Tian$^2$, Alexander Altland$^3$}
\affiliation{$^1$Department of Physics and Institute of Theoretical
Physics and Astrophysics, Xiamen University, Xiamen 361005, China\\
$^2$Institute for Advanced Study, Tsinghua University, Beijing 100084, China\\
$^3$Institut f{\"u}r Theoretische Physik, Universit{\"a}t zu K{\"o}ln,
K{\"o}ln 50937, Germany}

\date{\today}
\begin{abstract}

The quantum kicked rotor (QKR) driven by $d$ incommensurate
frequencies realizes the universality class of $d$-dimensional disordered
metals. For $d>3$, the system exhibits an Anderson metal-insulator transition
which has been observed within the framework of an atom optics realization.
However, the absence of genuine randomness in the QKR reflects in critical
phenomena beyond those of the Anderson universality class. Specifically, the
system shows strong sensitivity to the algebraic properties of its effective
Planck constant $\tilde h \equiv 4\pi /q$. For integer $q$, the system may
be in a globally integrable state, in a `super-metallic' configuration
characterized by diverging response coefficients, Anderson localized, metallic,
or exhibit transitions between these phases. We present numerical data for
different $q$-values and effective dimensionalities, with the focus being
on parameter configurations which may be accessible to experimental
investigations.
\end{abstract}
\maketitle

\section{\label{sec:introduction}Introduction}

The (quasiperiodic) quantum kicked rotor is a quantum particle on a unit
radius ring whose dynamics is described by the time dependent Hamiltonian
\begin{align}
\label{eq:QQKR}
\hat
H(t)=\frac{1}{2}(\tilde h \hat n)^2 + K\cos\hat\theta f_d(t)\sum_m \delta (t-m),
\end{align}
where $\hat \theta$ and $\hat n=-i\partial_\theta$ are coordinate and angular
momentum operator, respectively. The Hamiltonian $\hat H$ describes kicking of
the particle at unit time intervals and an amplitude depending on the angular
position. The quasiperiodic quantum kicked rotor given by Eq. \eqref{eq:QQKR}
differs from its more widely known sibling, the standard QKR~\cite{QKR79}, in
that the kicking strength itself, $\sim K f_d(t)$ is explicitly time dependent,
where the modulating function, $f_d(t)=\prod_{i=1}^{d-1}\cos (\omega_i t+ \phi_i)$
depends on $d-1$ incommensurate frequencies $\omega_i$. ($\phi_i$ are constant
phase offsets.) Much like that the standard QKR has been shown to lie in the
universality class of quasi-one dimensional disordered metals~\cite{Izrailev90,
Chirikov79, Fishman10, Fishman83, Fishman84, Altland10}, the quasiperiodic QKR
corresponds to a $d$-dimensional metal~\cite{Altland11}. (The mapping to a
$d$-dimensional effective system will be made explicit below.) The Anderson
localization phenomena characteristic for both one-dimensional~\cite{Raizen95}
and higher dimensional~\cite{Deland08, Deland10, Deland12} metallic systems have
been seen in cold atom experiments. Strikingly, a three-dimensional quasiperiodic
QKR has been experimentally shown to undergo an Anderson metal-insulator
transition upon variation of the kicking amplitude.

\begin{figure}[b]
\hskip-.5cm
\includegraphics[width=7.6cm]{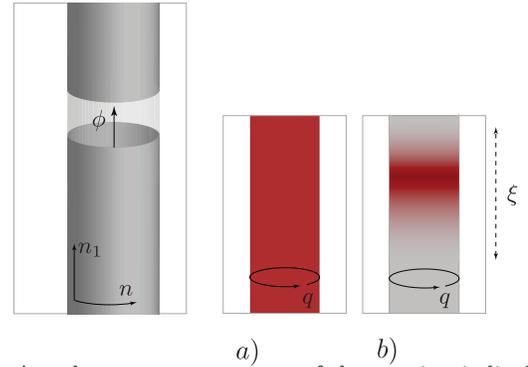}
\vskip-.5cm
\caption{Angular momentum space of the quasiperiodic QKR at a resonant value
$\tilde h = 4\pi p/q$. The system becomes effectively finite in $n$-direction
but remains infinite in its $d-1$ auxiliary dimensions ($d=2$ in the figure.)
Physical observables can be computed by probing the sensitivity to boundary
conditions in $n$-direction or, equivalently, to an Aharonov-Bohm flux, $\phi$
piercing the system. The ensuing physics then crucially depends on whether
wave functions are extended, a), or localized, b), in the auxiliary directions.
}
\label{fig:RotorResponse}
\end{figure}

The fact that the rotor is a deterministic chaotic, rather than a stochastic
disordered systems manifests itself in various anomalies emerging at specific
values of the global kicking strength, $K$, and Planck's constant $\tilde h$
(see Refs.~\cite{Altland10, Izrailev90, Fishman03a, Wimberger} for review on
anomalies of the standard rotor). Of particular interest are `quantum resonances'
arising at values $\tilde h/(4\pi)=p/q$, where $p,q$ are co-prime integers.
At these values, the Hamiltonian \eqref{eq:QQKR} commutes with translations
$\hat n \to \hat n + q$ in angular momentum space. The one-dimensional ($d=1$)
standard rotor then ceases to be Anderson localized and behaves like a finite
size metallic system of extension $q$ instead leading to a quadratic growth
of the rotor's energy at large times. (For $q$  larger than the localization
length $\xi$ of the system a crossover to localization  takes place.) In Refs.
\cite{Altland11, Tian13} we have analytically shown that in $d>1$ the same
mechanism may lead to novel type of quantum criticality, outside the Anderson
universality class. Basic features of this phenomenon can be understood by
observing that at the resonant values the rotor becomes effectively finite
in the $n$-coordinate, while it remains infinitely extended in the $d-1$
auxiliary dimensions associated to the additional driving frequencies
\cite{Altland11}. Upon compactification of the `unit cell' in $n$ direction,
the system assumes the topology of a $d$-dimensional cylinder, and physical
observables such as the expectation value of the rotor's energy, $E(t)\equiv
\left\langle \hat n(t)^2 \right\rangle$, can be computed by probing its
sensitivity to changes in the boundary conditions in the compact $n$-direction.
The behavior of the above expectation value, which in the metallic analogy
is the Fourier transform of the frequency dependent optical conductivity,
crucially depends on the localization properties in the infinitely extended
$d-1$ dimensions of the cylinder (cf. Fig.~\ref{fig:RotorResponse}.) In
dimensions $d\ge 3$, above the Anderson metal-insulator transition, wave
functions are extended, the system then resembles an ordinary metal, with
finite optical conductivity. However, below the Anderson transition, or
in low dimensions $d\le 3$, wave functions are localized, which means that
`transport' in $n$-direction is via a discrete spectrum of (localized) states.
In this phase the system has much in common with a `super-metallic' quantum
dot and the discreteness of its spectrum implies a diverging optical
conductivity. Somewhat counter-intuitively, this supermetallic conduction
behavior is rooted in strong Anderson localization in the transverse
directions.

In Ref.~\cite{Altland11}, the existence of a supermetallic phase in low
dimensions, and of a metal-supermetal transition in dimensions $d>3$ was
predicted on the basis of a field theoretic analysis. The purpose of the
present paper is to put these results to a numerical test. At the same time,
we will pay attention to anomalies arising at small values $q=1,2$ where the
system becomes integrable and instead of localization,quasiperiodic oscillatory
patterns is observed (cf. the left column of Tab.~\ref{tab:1} in which the main
observations of this paper are summarized). We have also identified anomalies
arising at $q=4$, where the integrability is partially restored and consequently
the generic picture breaks down and metallic regimes are absent (Tab.~\ref{tab:1}
right column.) The general conclusion will be that the adjustability of the two
principal parameters $(K,\tilde h)$ provides us with a spectrum of opportunities
to realize critical phenomena pertaining to the physics of integrability, chaos,
and localization. The physics addressed in the present paper should be well in
reach of current experiments~\cite{Deland08, Deland10, Deland12}.

\begin{table*}
\newcommand{\tabincell}[2]
{\begin{tabular}{@{}#1@{}}#2
\end{tabular}}
\centering
\caption{\label{tab:1}Summary of main results.}
\begin{tabular}{c||c|c||c|c|c||c|c|c}
\hline\hline
\multirow{2}{*}{parameter} & \multicolumn{2}{c||}{$q=1,2$} &
\multicolumn{3}{c||}{$q=3,5,6\dots$} & \multicolumn{3}{c}{$q=4$}\\
\cline{2-9}
&$\langle \hat n^2(t)\rangle
$ & phase & $\langle \hat n^2(t)\rangle
$ & phase & crossover time &$\langle \hat n^2(t)\rangle$
& phase & crossover time \\
\hline
$d=2$&\multirow{5}{*}{\tabincell{c}{quasiperiodic \\ oscillation}}
&\multirow{5}{*}{integrable}&\multirow{2}{*}{$\sim t^2$}&\multirow{2}{*}
{supermetal}&$t_\xi \sim K^2$&\multirow{4}{*}{$\sim t^2$}&\multirow{4}{*}
{supermetal}&\multirow{4}{*}{$t_\xi \sim K$}\\
\cline{1-1}
\cline{6-6}
$d=3$&&&&& $\ln t_\xi \sim K^2$&&&\\
\cline{1-1}
\cline{4-6}
\multirow{2}{*}{$d=4$}&&&$\sim t^2\; (K<K_c)$&supermetal& $t_\xi \sim
(K_c-K)^{-\alpha}$&&&\\
\cline{4-6}
&&&$\sim t\; (K\ge K_c)$ & metal & $\infty$&&&\\
\hline
\end{tabular}
\end{table*}
\vskip0.6cm

The rest of the paper is organized as follows: in
section~\ref{sec:Floquet}, we introduce the Floquet operator underlying
our analysis. In sections~\ref{sec:Integrable}, \ref{sec:MetalSupermetal},
and~\ref{sec:Anomalous},  we will simulate its dynamics to explore the
behavior at the smallest resonant values, $q=1,2$, `generic' resonant values
$q=3,5,\dots$, and the anomalous value $q=4$, respectively. We conclude in
section~\ref{sec:Discussion}.

\section{\label{sec:Floquet}Floquet operator}

Below, we will apply fast Fourier transform techniques to simulate the
quantum evolution of the initial state $|n\equiv 0\rangle$ at integer
times $t$ as $|\psi(t)\rangle = \prod_{s=1}^t \hat U'(s)|0\rangle$, where
\begin{eqnarray}\label{eq:2}
\hat U'(s)\equiv e^{-\frac{i\tilde h \hat n^2}{2}}
e^{-\frac{iK}{\tilde h}f_d(s)\cos\hat \theta},
\end{eqnarray}
is the Floquet operator. Using these states we numerically compute
the expectation value $E'(t)=\langle \hat n^2(t)\rangle =-\langle \psi(t)|
\partial_\theta^2 |\psi(t)\rangle$ to learn about the physical properties of
the system. The operator \eqref{eq:2} explicitly depends on the discrete time,
$s$, and in this non-autonomicity hides the effective dimensionality of the
system. Following ideas introduced in Refs.~\cite{Altland11, Casati89}, we
briefly review how the time dependence of $\hat U'$ may be eliminated at the
expense of  introducing $d-1$ additional dimensions. To this end, let us
interpret $|\theta_0 \equiv \theta, \theta_1, \cdots, \theta_{d-1}\rangle$
as a $d$-dimensional coordinate vector, comprising a `real' angular coordinate,
$\theta$, and a generalization of the parameters $\theta_{i\geq 1}$ entering
the definition of the kicking function, $f_d$, to `virtual' coordinates.
Corresponding to the `coordinate state', we have a $d$-dimensional angular
momentum state, $|n_0\equiv n, n_1,\cdots, n_{d-1} \rangle$, where
$\hat n_i\equiv -i\partial_{\theta_i}$ is conjugate to $\theta_i$. The gauge
transformed operator
\begin{eqnarray}\label{eq:3}
\hat U &\equiv& e^{-i(s+1)\sum_{i=1}^{d-1}\omega_i\hat n_i}\hat U'(s)
e^{is\sum_{i=1}^{d-1}\omega_i\hat n_i} \nonumber\\
&=& e^{-i\left(\frac{\tilde h \hat n^2}{2}+\sum_{i=1}^{d-1}
\omega_i\hat n_i\right)} e^{-\frac{iK}{\tilde h}
\prod_{i=0}^{d-1}\cos\hat\theta_i},
\end{eqnarray}
then turns out to be time-independent. It acts in the effectively
$d$-dimensional Hilbert space corresponding to the states above.

Physical observables are to be computed at a fixed value of the phases
$(\theta_1,\dots,\theta_{d-1})$, which means a trace over the conjugate
momenta. In the definition of our observables, $E(t)$, this trace is implicit.
In the following sections, we will explore the behavior of the expectation
value for various values of the parameters $q, K, d$. In doing so, we will
be met with a different types of behavior, where a saturation
$E(t)\stackrel{t\to\infty}{\longrightarrow}\mathrm{const.}$ indicates Anderson
localization, $E(t)\sim t$ is a characteristic for diffusive dynamics in the
angular momentum space, and  $E(t)\sim t^2$ for super-metallic behavior. In
cases with localization, the time $t\sim t_\xi$ at which saturation sets in
marks the localization time. Finally, persistent quasiperiodic fluctuations
in $E(t)$ are indicative of integrable dynamics.

In our simulations below, we will employ both representations, $\hat U$ and
$\hat U'$, and the expectation values $\left\langle \hat n^2(t)\right\rangle$
obtained in this way will be denoted $E(t)$ and $E'(t)$, respectively. The
gauge equivalence of the two representation implies $E(t)=E'(t)$.

\section{Integrable dyanamics at $q=1,2$}
\label{sec:Integrable}

\begin{figure}[b!]
\vskip-.2cm\hskip-.5cm
\includegraphics[width=8.6cm]{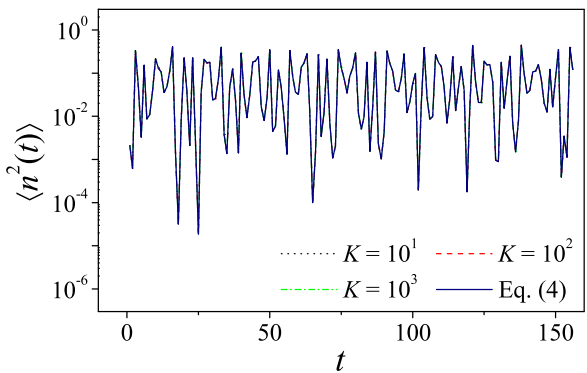}
\vskip-.5cm
\caption{Both simulations and analytic results -- in perfect agreement --
show that $\langle \hat n^2(t)\rangle$ (in unit of $\frac{1}{2}(K/\tilde h)^2$)
exhibits quasiperiodic oscillations.
}
\label{fig:q2d3}
\end{figure}

For $q=1,2$, the function  $E(t)$ exhibits quasiperiodic oscillations,
irrespective of the values of of $K$ and $d$. The origin of these oscillations
is the integrability of the rotor at $q=1,2$. Indeed, it is straightforward to
verify that
\begin{align}
	&\langle n|\prod_{s=1}^t \hat U'(s) |m\rangle=\crcr
	&\quad=\left\{
	\begin{array}{ll}
		J_{n-m}(\frac{K}{\tilde h}
\sum_{i=1}^t f_d(i)),&\; q=1,\crcr
(-)^{n-m\delta_{1,{\rm P}}} J_{n-m}
(\frac{K}{\tilde h} \sum_{s=1}^t (-)^{s+{\rm P}}f_d(s)),&\; q=2,
	\end{array}\right.\nonumber
\end{align}
$J_n(x)$ is the Bessel function and ${\rm P}$ is the parity of the (discrete)
time $t$: for even (odd) $t$ we have ${\rm P}=+1$ ($-1$), and  we are staying
in the un-gauged one-dimensional representation of the system. Using these
matrix elements we obtain
\begin{eqnarray}
\frac{\langle \hat n^2(t)\rangle}{\frac{1}{2}(K/\tilde h)^2}=
\Bigg\{\begin{array}{ll}
\left(\sum_{s=1}^t f_d(s)\right)^2,&\; q=1, \\
\left(\sum_{s=1}^t (-)^s f_d(s)\right)^2,&\; q=2,
\end{array}
\label{eq:4}
\end{eqnarray}
This shows that $E(t)/[\frac{1}{2}(K/\tilde h)^2]$ which collapses onto a
universal curve, independent of $K$, but dependent on $d$. Fig.~\ref{fig:q2d3}
compares simulations and the analytical result \eqref{eq:4} for $d=3$, $q=2$,
and parameters $(\omega_1, \phi_1)=2\pi((\sqrt{5}-1)/2, \sqrt{3}-1)$, $(\omega_2,
\phi_2)=2\pi(\sqrt{2}, \sqrt{11}-3)$. Analytical results and numerics are in
perfect agreement. The curves illustrate how the rotor's energy exhibits
quasiperiodic oscillations of rather small amplitude. The immobility of the
system in $n$-space effectively makes it as an insulator.

\section{Metal-supermetal transition at $q=3,5,6\dots$}
\label{sec:MetalSupermetal}

We now consider the value $q=3$, which defines the first configuration where
integrability is lost.  The resulting phenomenology crucially depends on the
effective dimensionality of the system, and we discuss various cases separately.
Numerically, we have found the system's behavior at $q=5,6,7\cdots$ is the
same as at $q=3$.

\subsection{\label{sec:supermetal}QKR as a supermetal at $d=2,3$}

\begin{figure}[h]
\vskip-.2cm\hskip-.5cm
\includegraphics[width=8.6cm]{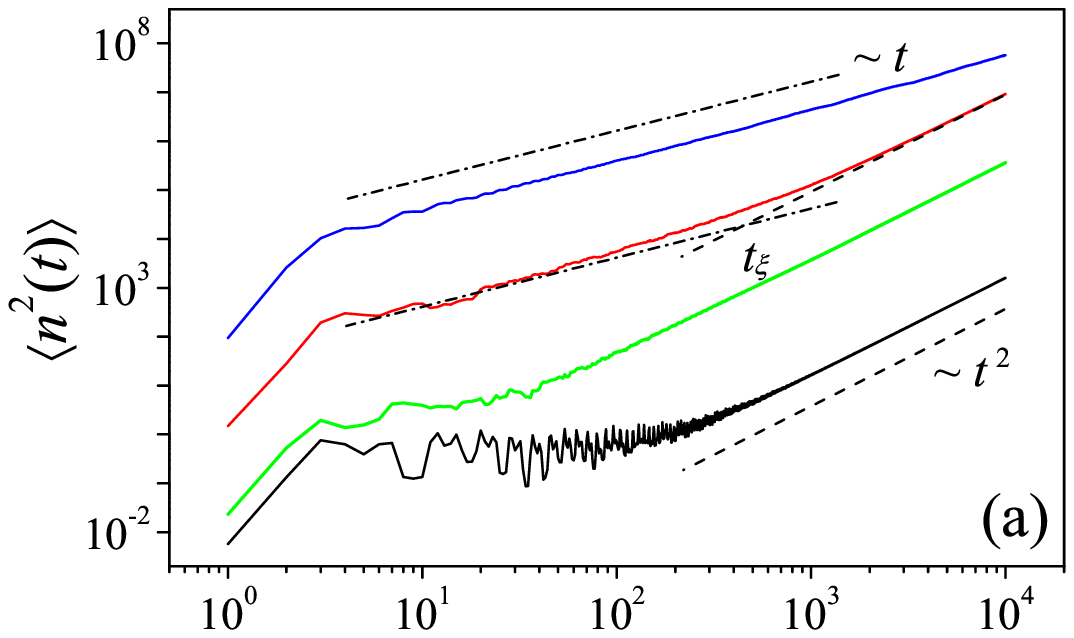}
\vskip-.65cm\hskip-.5cm
\includegraphics[width=8.6cm]{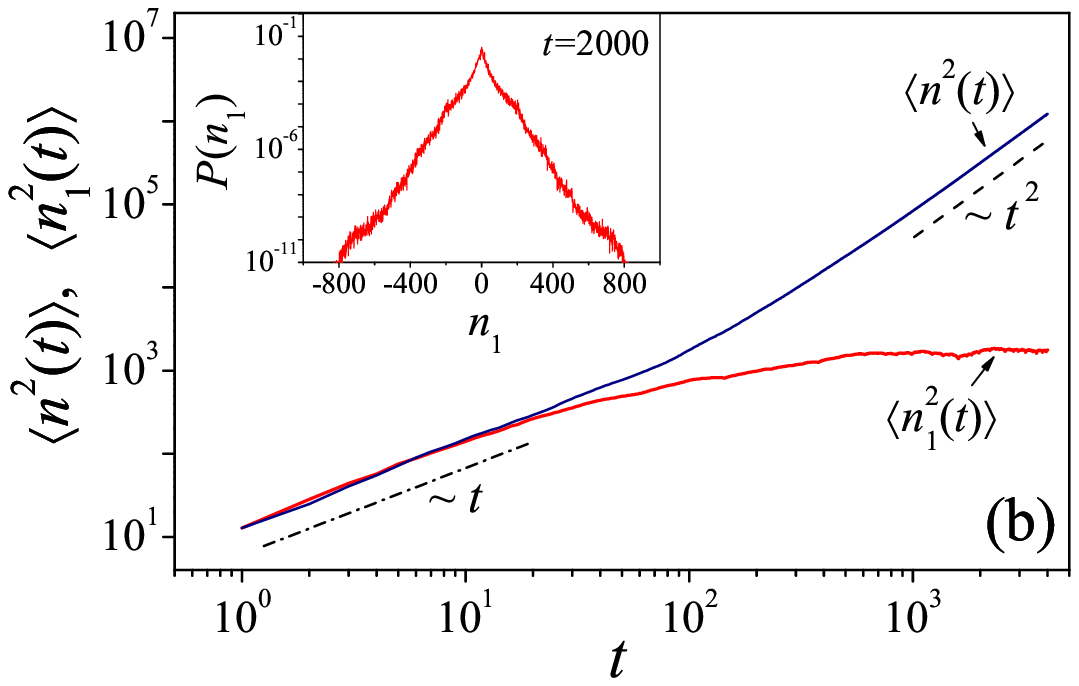}
\vskip-.5cm
\caption{(a) For $d=2,q=3$ the QKR exhibits a supermetallic energy growth,
$\langle \hat n^2(t)\rangle\sim t^2$,  at large times. From bottom to top,
the solid curves are for $K=4$, $8$, $64$, and $512$, respectively. (b) The
saturation of $\langle \hat n_1^2(t)\rangle$ and the supermetallic growth of
$\langle \hat n^2(t)\rangle$ simultaneously occur. $K=30$. Inset: quasi
$1$D Anderson localization in the $n_1$-direction.}
\label{fig:d2q3}
\end{figure}

To realize a $d=2$ dimensional system, we modulate the pulse amplitude with
one frequency $\omega_1$ ($d=2$) and simulate the dynamics (\ref{eq:2}) with
the parameters $(\omega_1, \phi_1)$ given above. Results for $E(t)$ are shown
in Fig.~\ref{fig:d2q3}(a), where the $\sim t^2$ asymptotic at large times
reflects supermetallic behavior. For large $K$ (e.g., $K=64$) the energy
growth displays a clear metal-supermetal crossover.

To better expose its origin, we simulate the $2$D dynamics in terms of $\hat
U$, Eq.~\eqref{eq:3}, and compare the expectation value $E(t) = \langle \hat
n^2(t)\rangle$ to the momentum dispersion in the virtual direction $\langle
\hat n^2_1(t)\rangle$. The results  shown in Fig.~\ref{fig:d2q3}(b) demonstrate
localization in the virtual $n_1$-direction and delocalization in the real
$n$-direction. It is also evident that the crossover  to supermetallic growth
and localization in the virtual direction takes place at the same time, $t_\xi(K)$.
The inset of Fig.~\ref{fig:d2q3}(b) explicitly shows the exponential decay
of a wave function amplitude projected onto the $n_1$-direction, denoted as
$P(n_1)$. These results indicate that the analytic predictions obtained for
large $q$ in Ref.~\cite{Altland11, Tian13} remain valid even for small $q$.

\begin{figure}[t!]
\vskip-.2cm \hskip-.5cm
\includegraphics[width=8.6cm]{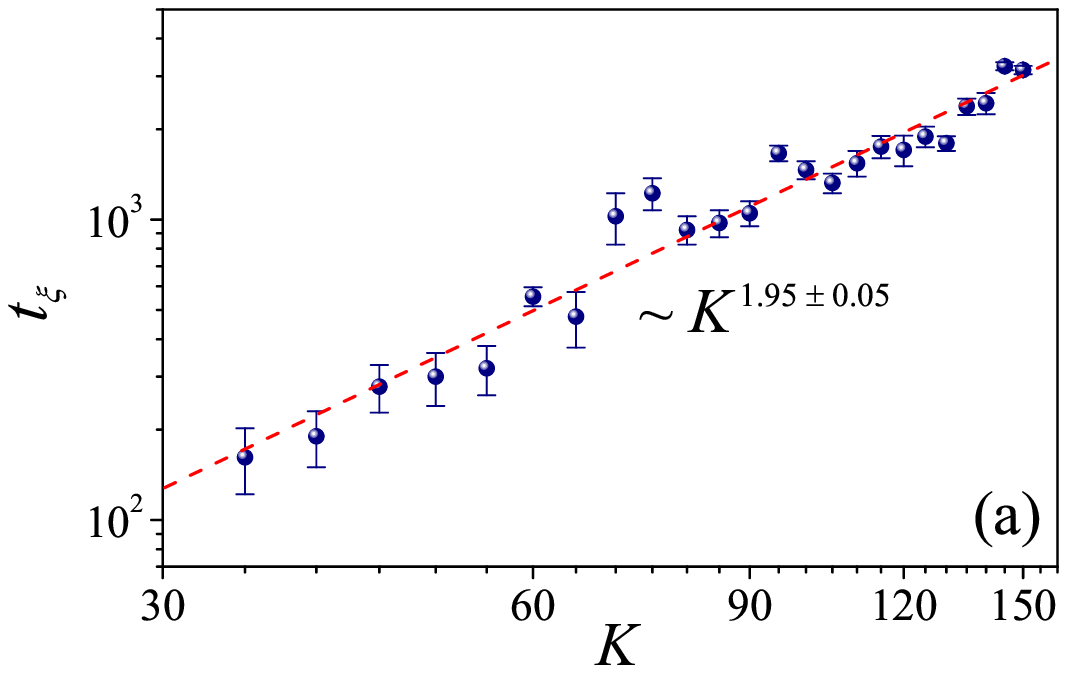}
\vskip-.7cm\hskip-.5cm
\includegraphics[width=8.6cm]{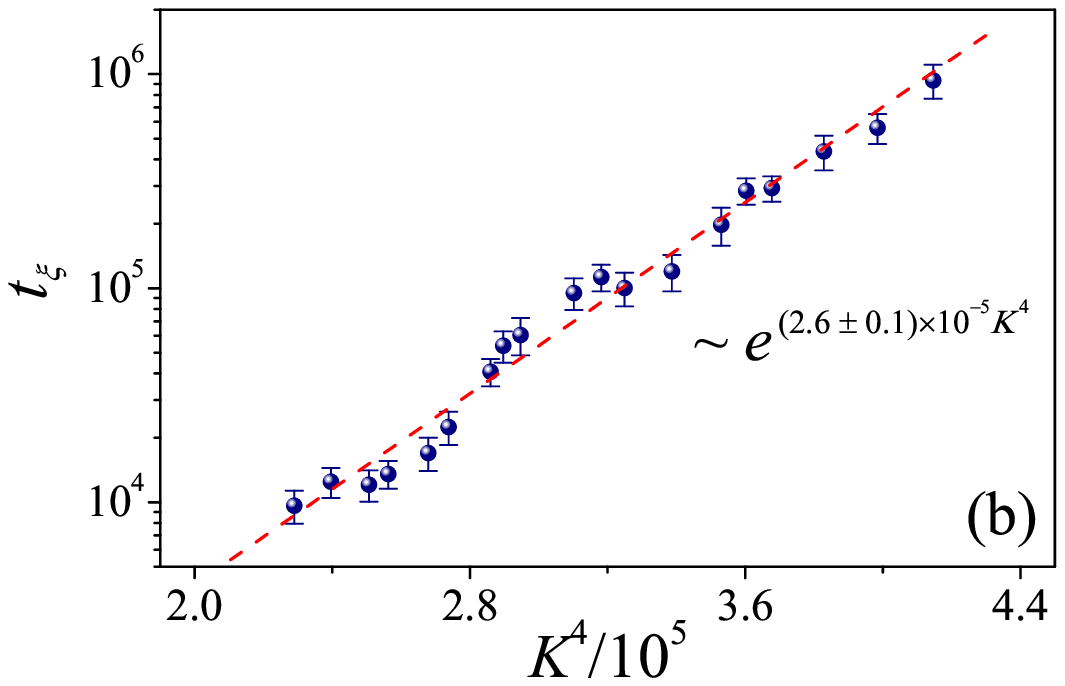}
\vskip-.5cm
\caption{The scaling behavior of $t_\xi$ for (a) $d=2$ and (b) $d=3$ at
$q=3$ in the system (\ref{eq:2}). The dashed lines are for the best linear
fitting results.}
\label{fig:d2q3tcr}
\end{figure}

We next discuss the scaling behavior of $t_\xi(K)$. To this end we extrapolate
the short and the long time power laws pertaining to the metallic (supermetallic)
growth to larger (smaller) times in $E'(t)$. In a double-logarithmic
representation, this produces two straight lines with a crossing point whose
time coordinate we identify with  $t_\xi$ (cf. Fig.~\ref{fig:d2q3}(a).) The
results of this analysis are shown in Fig.~\ref{fig:d2q3tcr}(a), and a power
law fit obtains $t_\xi\propto K^{1.95\pm 0.05}$. This is in excellent
agreement with the analytic prediction~\cite{Altland11, Tian13} $t_\xi
\propto D\stackrel{q,K\gg 1}{\sim} K^2$, where $D$ is the classical diffusion
coefficient. At small values of $K$ the diffusion constant becomes subject to
short time correlation corrections oscillatory in $K$, and this leads to the
growth of deviations off the $K^2$ asymptotic.

The above results show that
the behavior of $E(t)$ at $q=3$ is explained by the same physical mechanisms
as in the analytically studied $q\gg 1$ case: for short times, $t\ll t_\xi$,
the dynamics of wave packets in angular momentum space is diffusive. At the
corresponding frequency scales, $\omega\sim t^{-1}\gg t_\xi^{-1}\sim
\Delta_\xi$, where $\Delta_\xi$ is the spacing between adjacent localized
levels, the spectrum probed by the response function effectively looks
continuous, or metallic. In the long time regime, $t\gg t_\xi$, wave packets
are localized, and the conjugate frequencies $\omega\ll \Delta_\xi$ are small
enough to probe individual localized states. A straightforward
analysis~\cite{Altland11,Tian13} shows that this leads to a divergent optical
conductivity, or linear scaling $\sim t$ of the function $E(t)$.

In the case of $d=3$, simulations of the rotor driven by two frequencies
$\omega_{1,2}$ show that $E'(t)$ crosses over from linear to quadratic increase
at time $\sim t_\xi$, as in $d=2$. However, as shown in Fig.~\ref{fig:d2q3tcr}
(b), $t_\xi$ now grows exponentially in $K^4 \sim D^2$. Again we see that at
small values of $K$ short time correlation corrections oscillatory in $K$ leads
to the growth of deviations off the $K^4$ asymptotic. This scaling reflects the
exponential dependence of the localization length on the square of the diffusion
coefficient characteristic for effectively $2$-dimensional (localization is in
the $d-1$ dimensional virtual space) disordered systems~\cite{Altland11}. This
is a manifestation of unitary Anderson localization in the $2$-dimensional
virtual space, as expected by the field theoretic analysis~\cite{Altland10,
Altland11, Tian13}.

\begin{figure}[h!]
\vskip-.2cm \hskip-.5cm
\includegraphics[width=8.6cm]{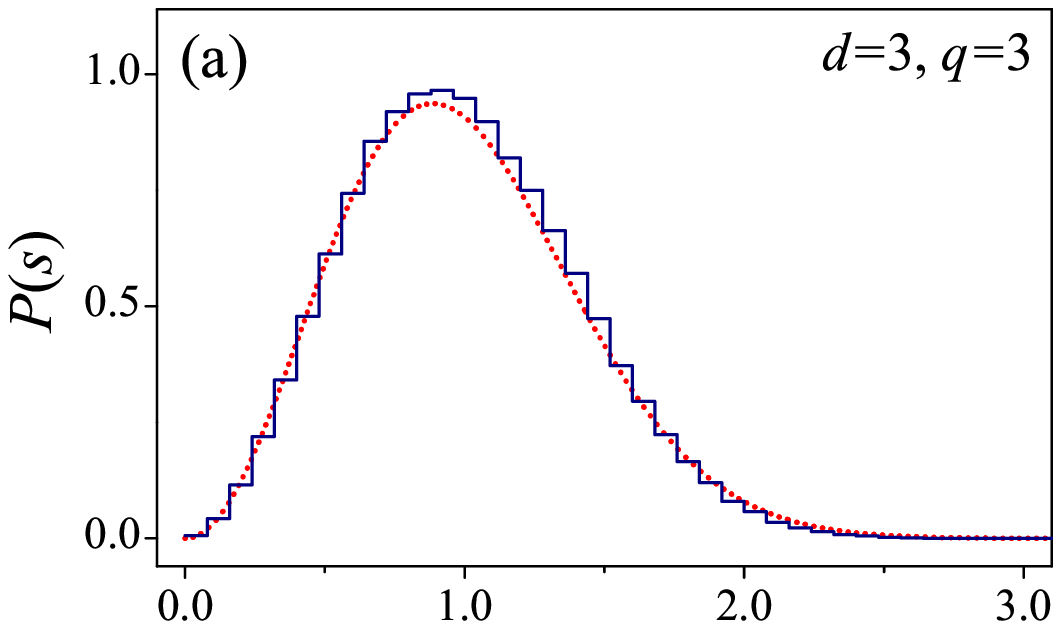}
\vskip-.5cm\hskip-.5cm
\includegraphics[width=8.6cm]{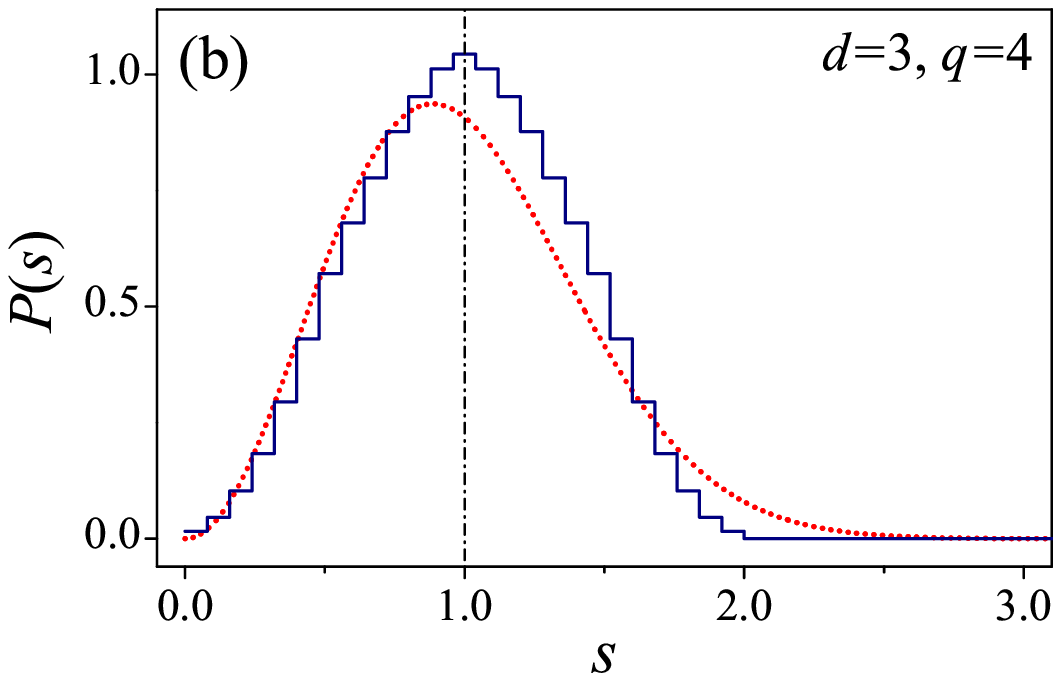}
\vskip-.5cm
\caption{The level spacing distribution (histogram) for $q=3$ (a) and $q=4$
(b) in the three-dimensional system (\ref{eq:3}) with $K=80$. The red
dotted lines in both panel represent the Wigner surmise for CUE. Note
that in (b) $P(s)$ is symmetric with respect to $s=1$. The parameters $\omega_{1,2}/(2\pi)$
are approximated by $13/21$ and $23/17$, respectively.}
\label{fig:spectrum}
\end{figure}

Indeed, the $q$-periodicity in $n$-direction introduces an
Aharonov-Bohm flux, $\phi$, namely the Bloch momentum piercing the system
(cf. Fig.~\ref{fig:RotorResponse}) which effectively breaks the time-reversal
symmetry of quantum dynamics within a unit cell. To confirm this symmetry we
further perform a study of spectrum statistics. To this end we approximate
$\omega_{1,2} /(2\pi)$ by rational number and compactify the unit cell in
$n_{1,2}$-direction. For the ensuing torus we perform numerical diagonalization
and find the quasienergy spectrum for fixed Bloch momentum $\phi$. Then, by
scanning $\phi$ we obtain a large ensemble. This allows us to compute the level
spacing distribution, denoted as $P(s)$. As exemplified in Fig.~\ref{fig:spectrum}
(a), the results are in excellent agreement with the Wigner surmise for the
circular unitary ensemble (CUE). (We recall for the standard one-dimensional
QKR, it has been analytically shown that the unitary symmetry leads to a
simple, universal linear to quadratic crossover in the rotor's energy
growth~\cite{Altland10}.)

\subsection{\label{sec:transition} Metal-supermetal transition at $d=4$}

\begin{figure}[t]
\vskip-.3cm \hskip-.5cm \includegraphics[width=8.6cm]{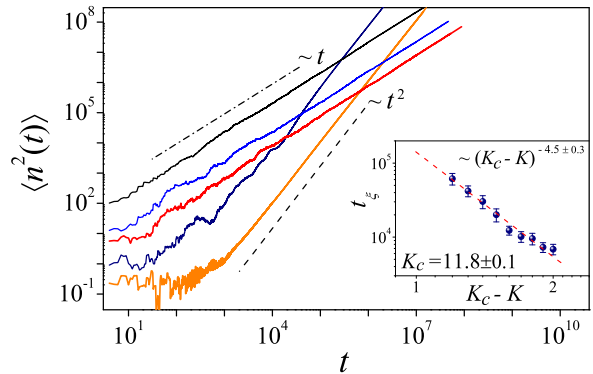}\vskip-.5cm
\caption{For $d=4$ and $q=3$, the QKR displays a metal-supermetal transition
as $K$ decreases. From bottom to top at the left side, the solid curves are
for $K=4,8,20, 30$, and $80$, respectively. Inset: the crossover time $t_\xi$
exhibits criticality.}
\label{fig:d4q3}
\end{figure}

Moving up in dimensionality, we introduced a third frequency/phase pair
$(\omega_3, \phi_3)=2\pi((\sqrt{7}+1)/2, \sqrt{17}-4)$ to simulate the system
at $d=4$. Fig.~\ref{fig:d4q3} shows results of $E(t)$ for different values of
$K$. Our simulations indicate that at $K_c=11.8\pm 0.1$ the long-time behavior
undergoes a transition from quadratic to linear large time asymptotics. This
is the Anderson transition separating an Anderson localized from a metallic
phase in three dimensional virtual space. We have found that the localization
time for small deviations of $K$ off the critical values scales as $t_\xi \sim
(K_c-K)^{-\alpha}$  (Fig.~\ref{fig:d4q3} inset) with a critical exponent
$\alpha =4.5\pm 0.3$. These observations are again in agreement with the large
$q$ results obtained in Ref.~\cite{Altland11}.

Unlike in $d=2,3$, simulations of the $4$-dimensional operator~\eqref{eq:3},
i.e. of the function $E(t)$, are difficult. However, the observed value of
$K_c$, and the value of the critical exponent $\alpha$ can both be understood
from scaling arguments: Anderson localization in virtual space leads to a
frequency dependent renormalization of the diffusion coefficient, $D\to D(\omega)$,
where $\omega$ is Fourier conjugate to the observation time. Similar to discussions
in Sec.~\ref{sec:supermetal}, the periodicity in $n$-direction renders Anderson
transition in the $(d-1)$-dimensional virtual space of unitary type. Correspondingly,
by using the standard renormalization group analysis the leading (localization)
correction is given by $D(\omega)\approx D[1-\frac{1}{2\pi q^2 D}\int \frac{d^{d-1}
\phi}{(2\pi)^{d-1}} (-i\omega+D\phi^2)^{-1}]$. For $d\geq 3$ the integral suffers
ultraviolet divergence and requires a short distance cutoff $\sim {\cal O}(K/
\tilde h)$.
Then, a rough estimate for the onset of strong localization follows from the
equality of the constant classical contribution to the quantum correction, i.e.
from the condition $D(\omega=0)\approx 0$. Doing the integral, we obtain the
equivalent condition (for $d=4$)
\begin{equation}\label{eq:5}
(4q^2\pi^3)^{1/5}\frac{K_c}{8\tilde h}={\cal O}(1),
\end{equation}
which is well satisfied by the observed value $K_c\approx 11.8$ (at which the
left hand side of Eq.~(\ref{eq:5}) equals $1.4$.)

Beyond perturbation theory~\cite{Deland09, Altland11} the diffusion coefficient
$D(\omega)$ scales as $D(\omega)=\omega^{\frac{1}{3}}f((K-K_c)\omega^{-\frac{1}
{3\nu}})$, where $f(x)$ is some scaling function, and $\nu>0$ is the localization
length critical exponent, i.e. $\xi \sim (K_c-K)^{-\nu}$. Noting that $\omega\sim
t^{-1}$, we conclude that in the virtual space the wave packet expansion saturates
at large times when $(K_c-K)t^{\frac{1}{3\nu}}\gg 1$. This implies that in
the supermetallic phase the metal-supermetal crossover occurs at $t_\xi\sim
(K_c-K)^{-3\nu}$, i.e. we have arrived at the identification $\alpha=3\nu$.
Our simulations predict that $1.4 \leq \nu \leq 1.6$ consistent with
general results for the $3$-dimensional Anderson transition of unitary
type~\cite{Slevin97}.

The above results for $K_c$ and $\alpha$ corroborates the view that the
phase transition observed at $q=3$ is in the universality class of the
Anderson metal-insulator transition. Below the critical value $K=K_c$, the
system effectively behaves as a finite system of extension $q\xi^3$ and
finite size quantization of energy levels then is responsible for the
supermetallic scaling of response coefficients.

\section{Anomalous supermetallic behavior at
$q=4$}
\label{sec:Anomalous}

\begin{figure}[tt]
\vskip-.3cm \hskip-.5cm \includegraphics[width=8.6cm,height=6.4cm]{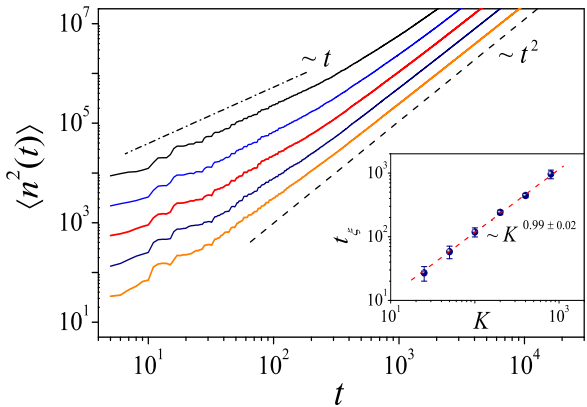}
\vskip-.4cm
\caption{The anomalous energy growth for $q=4$ with $d=3$. In the main panel,
the solid curves (from bottom to top) are for $K=25, 50, 100, 200$, and $400$,
respectively. The insert shows $t_\xi$ as a function of $K$. }
\label{fig:q4}
\end{figure}

Numerical experiments further show that for larger values of $q(=5,6,7,\dots)$
the QKR behaves in the same way as the $q=3$ case. This suggests that the
unconventional quantum criticality occurs for generic $q$. This notwithstanding,
anomalous behavior is observed for $q=4$: Regardless of the dimension, $d$,
the energy growth exhibits a linear-quadratic crossover with the crossover
time $t_\xi \sim K$. (See Fig.~(\ref{fig:q4}) as exemplified by the
case of $d=3$).

To understand why unusual things happen at this $q$ value, notice that in the
QKR context, the kinetic energy operator $\exp(-i\tilde h \frac{\hat n^2}{2})$
plays the role of a stochastic scattering operator, much like a random real
space potential in conventional Anderson localization. Our so far analysis
presumes that this operator does not exhibit any regular structure throughout
the unit cell, $n=0,\dots, q$. However, for $q=4$, this operator is translationally
invariant in $2$ and the unit cell, $\{0,1,2,3\}$, splits into two replicated
sub-cells $\{0,1\}$ and $\{2,3\}$. Most interestingly, this reduction renders
the rotor similar to its genuine $2$-periodic sibling: the only difference is
that in the former (latter) the factor $\exp(-i\tilde h \frac{n^2}{2})$ takes
the value of $-i$ ($-1$) for odd $n$. On general grounds, we expect integrability
to be (partially) restored. Indeed, we find that the level spacing distribution
is dramatically different from the $q=3$ case: strikingly, it is symmetric with
respect to $s=1$ and only for small $s$ it follows the Wigner surmise of CUE
type (see Fig.~\ref{fig:spectrum} (b)).

Moreover, our  numerical analysis for $q=4$
shows that an initial regime of diffusion -- a manifestation of stochasticity --
is followed by a strong tendency to localize already at times $t>K$ parametrically
shorter than in the generic case (cf. Fig.~\ref{fig:q4}.) While we do not fully
understand the origin of this behavior, it appears to be outside the standard
Anderson universality class. In addition, it is interesting to notice that at
$q=4$ no localization-delocalization transition is observed. We believe that
this is intrinsic to the partial restoration of integrability. Further research
is required to understand these phenomena and to explore if there exist any other
anomalous $q$ values.


\section{Discussion}
\label{sec:Discussion}

In this paper we have numerically explored the QKR driven by $d-1$ incommensurate
frequencies and at resonant values of Planck's constant $\tilde h=4\pi/q$. Compared
to the standard rotor, the presence of additional driving frequencies, and the fine
tuning of Planck's constant provide the option to realized qualitatively novel types
of quantum criticality. We have seen that, depending on the value of $q$, the system
may be integrable at $q=1,2$, be in the Anderson universality class on a circumference
$q$ cylinder of dimensionality $d$ ($q=3,5,6,\dots$), or in an anomalously localized
regime $(q=4)$. The option to change the universality class of the system by a well
defined change of a single control parameter provides us with a high-quality test bed
of our understanding of Anderson type quantum criticality. It stands to reason that
the configurations explored in this paper, $d=2,3,4$ and $q=1,2,3,4$ are within the
reach of state-of-art atom-optics setups~\cite{Deland08, Deland10, Deland12, Deland09,
Phillips06, Steinberg07}. In current experiments the expansion of atomic clouds can
be observed over several hundred kicks~\cite{Deland08, Deland10, Deland12, Deland09}
and a quantitative comparison to our results should be possible.

\section*{Acknowledgements}

Discussions with D. Delande, S. Fishman, J. C. Garreau, and I. Guarneri
are gratefully acknowledged. This work is supported by the NSFC (Grant
Nos. 11275159, 11335006, and 11174174), the Tsinghua University ISRP
(No. 2011Z02151), and the Sonderforschungsbereich TR12 of the Deutsche
Forschungsgemeinschaft.

\end{document}